# VIDEO CONTENTS PRIOR STORING SERVER FOR OPTICAL ACCESS NETWORK

Tony Tsang

Department of Computer Science, Chu Hai College of Higher Education, Hong Kong

*ABSTRACT*

*One of the most important multimedia applications is Internet protocol TV (IPTV) for next-generation networks. IPTV provides triple-play services that require high-speed access networks with the functions of multicasting and quality of service (QoS) guarantees. Among optical access networks, Ethernet passive optical networks (EPONs) are regarded as among the best solutions to meet higher bandwidth demands. In this paper, we propose a new architecture for multicasting live IPTV traffic in optical access network. The proposed mechanism involves assigning a unique logical link identifier to each IPTV channel. To manage multicasting, a prior storing server in the optical line terminal (OLT) and in each optical network unit (ONU) is constructed. In this work, we propose a partial prior storing strategy that considers the changes in the popularity of the video content segments over time and the access patterns of the users to compute the utility of the objects in the prior storage. We also propose to partition the prior storage to avoid the eviction of the popular objects (those not accessed frequently) by the unpopular ones which are accessed with higher frequency. The popularity distribution and ageing of popularity are measured from two online datasets and use the parameters in simulations. Simulation results show that our proposed architecture can improve the system performance and QoS parameters in terms of packet delay, jitter and packet loss.*

## 1.INTRODUCTION

The fast advance of optical access networking technologies makes various services through the IP network feasible. Especially, Internet Protocol Television (IPTV) has been a killer application of Internet services and provides new revenue opportunities for Internet service providers. IPTV provides digital television services including live television, time shifted television (TSTV), and video on demand (VoD), using the Internet protocol suite over the packet-switched and optical networks, such as the Internet. Generally, live broadcast TV channels are encoded and delivered in IP multicast streams, whereas TSTV and VoD services are provided in unicast streams. The number of video streams sent simultaneously by the head-end office to the consumer varies by the internet networks, but it is strictly restricted due to bandwidth limitation. A typical architecture of IPTV service system is depicted in this paper.

The key challenge in streaming media application is how to reduce data delivery latency, networks bandwidth and disk I/O bandwidth requirement in order to support more concurrent consumers [1]. An efficient approach is to deploy local video content servers close to the consumers at the access network. The distributed local video content servers allow operators to economically alleviate the inherent storage and network bandwidth limitation, proportionally distribute the subscriber load and service demand. The capacity of local video content server is also finite, however, an efficient strategy for improving quality of service (QoS) by storing popular video content that are likely to be used in the near future is strongly recommended. In this paper, therefore, an efficient prior storing management scheme is investigated for distributed IPTV local video content services. Video Content prior storage is a fundamental strategy for improving the performance and quality of service perceived by consumers. Furthermore, only several percentage point improvement in hit ratio by a good prior storing management is almost equivalent effect to a several-fold increase in storage sizes. Therefore a design of an efficient prior storing management algorithm is very important to enhance the performance of prior storing systems effectively. The popularity distribution of video content has a direct impact on the design of efficient prior storing strategy. Previous studies on VoD services frequently have used Zipf distribution to model the popularity of video object in their systems [2], [3].

In a typical VOD system, the probability of accessing a video (or content segment of a video) increases with its popularity. Some prior storing strategies have been proposed in the literature taking into account the popularity of the video and not just the recently of its access (which is taken care of by the Least Recently Used (LRU) scheme) or the frequency of its access (which is taken care of by the Least Frequently Used (LFU) scheme) [4]. However, it was rightly observed that most of the schemes assume that the popularity of the objects is known a prior and do not consider the change in popularity over time. Additionally, the user in a VOD system has the option to watch a video any number of times and can also decide to play only some parts of the video. Considering this behaviour, some parts of any video have higher popularity than the others can assume [5]. Prior storing these popular content segments would lead to a greater reduction in the bandwidth. Even if the video is partitioned into a fixed number of content segments, tracking the changes in the popularity of the content segments is an important requirement to design effective partial prior storing schemes. It was also observed that the popularity of the videos (or content segments) can decrease rapidly with time [4]. While it can take some time for a movie, since its induction into the system to become popular, the popularity can also decrease with time. Any partial prior storing scheme based on popularity should consider the volatile nature of popularity to increase the effectiveness of the prior storage.

Content Segment-based prior storing that does not track changes in popularity can lead to diminishing returns over time.

This work is motivated by the observation that neither a frequency-based technique like LFU nor a recently-based technique like LRU can give satisfactory performance for objects with volatile popularity (likelihood of future accesses) [6]. While the existing partial prior storing schemes fare better than full object storing schemes, the volatile nature of popularity and the viewing pattern should be considered to increase the effectiveness of the prior storage in a VOD system. In this work, we consider both the frequency and the recently of access for content segment prior storing and propose a method to dynamically compute the utility of a content segment stored in the prior storage with changes in the viewing pattern over time. We also observe that the objects in a VOD system have a wide variation in popularity over time. To protect the previously popular objects from being evicted by objects that see a sudden popularity, we propose to partition the prior store into two level prior storages. Each partitioned prior storing uses a different function to update the utility of content segments present in its storage. We also determine the popularity distribution and ageing in popularity based on user ratings from two online datasets.

The rest of the paper is organized as follows. In Section II, we discuss the proposed prior storing scheme along with the motivation behind the work. Section III presents the data from an analysis of Quality of Services from data sources and results from simulation of the proposed prior storing scheme. Section IV concludes the paper.

## 2. PROPOSED ARCHITECTURE AND OPERATIONS

### 2.1. Logical Link Identifier Operation

Access network devices attached to the Passive Optical Network (PON) medium implement the logical topology emulation (LTE) function based on its configuration and emulate either a shared medium or a point-to-point (PtP) medium [7,8]. IEEE Std 802.3 defines the LTE function in the residence gateways, which is below the MAC sub-layer, to hold the existing Ethernet MAC operation. The task of this function relies on the tagging of Ethernet frames with unique Logical Link Identifiers (LLIDs) for each ONU. The ONU LLID is assigned by the auto discovery. Within a certain time gap, each ONU needs to register in the OLT, and the OLT will send the REGISTER message to set up the ONU LLID for each ONU. When the OLT retrieve the REGISTER_ACK message, the OLT concurrently knows the ONU MAC and its LLID. Each ONU can have more than one LLID. The 16-bit LLID field placed in the preamble at the beginning of each frame has a mode bit and a 15-bit LLID. The mode bit shows the emulation mode, with 1 representing the shared medium emulation and 0 indicating the Point To Point emulation. The optical access network system works by verifying the LLID at the ONUs and the OLT. On the ONU side, filtering rules are defined to accept/reject the frames. The content frame can be accepted if the bit mode is 0 and the content frame LLID matches the ONU LLID, or if the bit mode is 1 without concern for the LLID value or if the frame LLID is broadcast/multicast LLID.

Figure 1 shows the proposed optical access network topology, which includes an OLT, a number of ONUs, and a 1:N passive splitter / combiner. To support the proposed architecture, the OLT and ONU hardware is redesigned. To multicast the IPTV channels, the Channel Logical Link Identifier (CLLID) is defined; it is a unique LLID that is assigned to each requested channel. The CLLID is independent of the ONU LLID. The mechanism is designed to assign the CLLID to the ONUs that request the channel. In this way, the OLT can assign more than one CLLID to one ONU. Consequently, the frames with the desired CLLID will be accepted by the ONU. With the 15-bit LLID, the OLT can generate 32,767 unique LLIDs. When the ONU requests/leaves a channel, the OLT assigns/removes the CLLID to/from the ONU. To manage the CLLID operations, a table is constructed in each ONU and OLT maintained in the residential gateway. The ONU table is composed of the channel_name, channel_LLID, user_MAC address, and user_IP address fields, whereas the OLT table contains the channel_LLID, channel_name, and ONU_LLID.

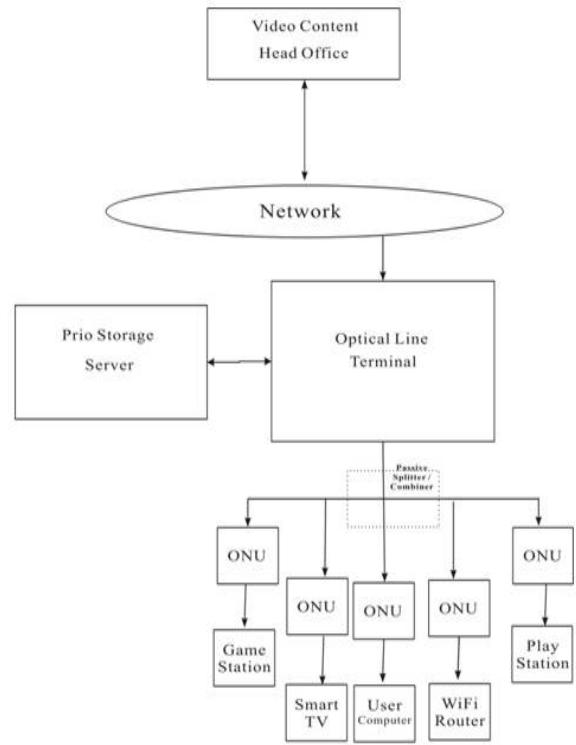

Figure 1. Proposed Optical Access network Topology

**2.2. Operation Functions for Bi-Level Prior Storing Scheme**

**2.2.1. Assumptions**

We assume a hierarchical IPTV network (operated by a single telecom provider) for the delivery of VOD services in which the video objects are located at the Video Contents Head Office and the Optical Network Units (ONUs) are connected to Optical Line Terminal (OLT). The network is supported by multiple levels of prior storages at the OLTs, the ONUs and the Input/ Outputs. We do not assume any cooperation among the prior storages in this work. We assume that video objects are partitioned into fixed size content segments. Based on conversations with system engineers at Laboratory, we found that fixed contents segmentation with a moderate size is preferred over adaptive contents segmentation owing to memory management issues. One problem with fixed contents segmentation is that the contents segment boundaries may not align with those of heavily-viewed portions, leading to prior storing inefficiency. To avoid this problem, we measure the utility of a content segment based on the number of bytes played from it, thereby capturing the utility of prior storing the contents segment. We assume that different contents segments of the same video might differ in their popularity. This is a reasonable assumption when the user has already watched the video earlier and is determined to watch some parts it again [9]. Changes in the popularity of a video at the segment level are also seen when the user can seek to watch different parts of the video (termed skipped viewing) so that some content segments of the

74

International Journal of Computer Networks & Communications (IJCNC) Vol.7, No.2, March 2015

video tend to have a higher popularity than the others. In this work, we use a content segment as a basic unit of a video object for prior storing and replacement. Henceforth, we use the words 'content segment' and 'object' in a prior storage interchangeably. Following [6], a large piece payload is used as the basic unit of a video and a content segment can be viewed as a collection of payloads.

### 2.2.2. Prior Storage Partitioning Scheme

In OLT-based prior storage, objects that were popular (but not often watched) in recent times, tend to get evicted from the prior storage. This can lead to the eviction of popular objects when the temporal distribution of the accesses to an object (content segment) is not uniform. Similarly, frequency-based prior storing schemes (LFU) do not perform well when the object pool is dynamic and the popularity of the objects in a prior store decreases with time (termed ageing) [6]. Popularity of an object in a prior storage increases with time if the number of requests made for that object increases. Typically, in a VOD system, the popularity of some objects could decrease with time if the viewers do not see the video often. Thus, it is necessary to consider the last time when the object was accessed along with its popularity when making a prior storage replacement. The utility of the objects already in a prior storage should consider both the frequency and recently of accesses to account for the temporal changes in popularity. Typically, a simple popularity-based prior storage scheme considers the ranking of objects before prior storage and does not adapt to the ageing popularity. Prior storage schemes that capture the ageing popularity should compute the utility for objects in the prior storage continuously, based on both recently and frequency of the accesses. Since this computation becomes complex with an increasing number of contents segments, the objects in the prior storage should not be evicted too soon before the computation is effective in reflecting the temporal changes in popularity. New objects with a high popularity inserted in the prior storage should not evict older objects (with relatively lower popularity) whose utility reflects the temporal distribution of accesses. Considering the aforementioned issues with prior storage schemes that ignore the volatile nature of popularity, we propose to divide a prior storage into two parts. *Prior Storage1* (primary prior storage) is used to prior storage a segment that is accessed for the first time (on a typical prior storage miss) and *Prior Storage2* (secondary prior storage) is used to store segments whose utility is already determined to be high (i.e., popular over a longer period of time). We use independent functions to determine the utility of segments in these two partitions because the eviction of contents segments from them should be handled differently. A segment that is accessed for the first time (or which is not in the prior storage) is prior stored in the primary prior storage. The utility of each segment in the primary prior storage is updated based on the recently and frequency of accesses made to it. Once the utility of a content segment crosses a threshold (a user-defined parameter), it is moved to the secondary prior storage. In this way, all contents segments in *Prior storage2* have a larger utility, determined over a longer period of time, when compared to the short-term popularity of segments in Prior storage1. The replacement of contents segments in both the partitions is handled independently according to the smallest-utility-first policy.

### 2.2.3. Dynamic Bi-Level Prior storage Scheme

A request is identified by an object ID, starting a large piece payload ID, and the time at which the request arrives. Whenever a request for an object is made, the object is fetched in a content segment-wise fashion. Depending on the state of contents segments (whether prior stored or not), our prior storage algorithm makes a decision as explained below.

Prior storage in *Prior Storage1*: If a requested content segment is neither present in *Prior storage1* nor in *Prior storage2*, then it will be prior stored in *Prior Storage1*. That is, *Prior*

*storage1* is used to store contents segments whose utility is yet to be determined. Clearly, our approach is aggressive for contents segments whose utility is yet to be determined. Subsequently, as the number of requests for the content segment increases, we determine its utility based on the average length of the content segment played and the recently of the request. Contents Segments in *Prior Storage1* are evicted based on the utility value determined by a utility function. Since the content segment is inducted into the prior storage without any prior information about its popularity, the utility must be proportional to the average number of large piece payload played in the content segment, and inversely proportional to the total number of payloads in the content segment and its last access time.

We determine the utility of contents segments in *Prior Storage1* as follows:

$$Utility_1 = \frac{N_{largepayloadplayed} * F_{recently}}{N_{largepayloadincontentSegment} + N_{request}} \qquad (1)$$

where

$$F_{recently} = \frac{1}{1 + (\frac{T_{current} - T_{lastaccessed}}{\beta})} \qquad (2)$$

Here, $N_{request}$ denotes the number of requests,

$N_{large\ payload}$ played denotes the number of large payload played, $N_{large\ payload}$ in content segment denotes the number of payload in a content segment, *Frecently* denotes the recently factor, $T_{current}$ denotes the time when the request arrives, and $T_{last}$ accessed denotes the time when the content segment was last accessed.

The utility value based on Eq.(I) decreases as the contents segmentages. This ensures that older segments are evicted from the prior storage. (3 is used to smoothen the recently factor, *Frequency*, based on the last access time, *Tlast* accessed. This ensures that the utility of the segments does not change by a large amount in a short duration, which otherwise leads to a false prediction of the popularity of contents segments.

While the content segment with the least utility is evicted, we do not evict a content segment being played to avoid a prior storage miss. If all the content segments in *Prior Storage1* are being played and a new request arrives for the content segment currently not in the prior storage, then the new request is dropped as there is no content segment to replace. When a content segment already in *Prior Storage1* is requested, the data related to that content segment such as, *Nrequest*. $N_{large\ payload}$ played, and *Tlast* accessed are updated.

Promotion to *Prior storage2*: For each request made for a content segment in *Prior Storage1*, its utility is checked; if this crosses a predefined threshold (*THRESHOLD*), then the content segment is marked for promotion to *Prior storage2*. The marked content segment is moved to *Prior storage2*. If necessary, another content segment in *Prior storage2* is replaced based on the utility values. *Prior storage2* is used to keep contents segments with a utility value that is greater than the *THRESHOLD*.

A different function is used to compute the utility of contents segments in *Prior storage2* because the popularity of contents segments is already determined before promotion. In *Prior storage2*,

International Journal of Computer Networks & Communications (IJCNC) Vol.7, No.2, March 2015

instead of using *Frecently* while calculating the utility, we determine the probability of the next request for the content segment. This gives a better prediction for the popularity of a content segment based on the inter-arrival time, $1/\lambda$, of requests for that content segment. As the popularity of a content segment decreases, the inter-arrival time between the requests for the content segment increases. The utility function for *Prior storage2* is given by

$$Utility_2 = \frac{N_{largepayloadplayed} * P_{nextrequest}}{N_{largepayloadinContentSegment} * N_{request}} \quad (3)$$

Here, $P_{next}$ request denotes the probability of the next request, given by

$$P_{nextrequest} = \frac{\frac{1}{\lambda}}{max\{\frac{1}{\lambda}, T_{sincelastrequest}\}} \quad (4)$$

where $T_{since\ last\ request}$ denotes the time since the last request.

Furthermore, to determine the correct utility of a content segment in a prior storage, we propose to keep each content segment in *Prior Storage1* for a minimum time that is proportional to the content segment size

**Algorithm 1** Dynamic Bi-Level Prior Storing Scheme:

**for each** request for an object; construct request for corresponding *Content Segment*,
**If** *Content Segment* is Prior Stored in *Prior Storage2*
Serve the *Content Segment* and update utility of the Segment.
**elseif** *Content Segment* is prior stored in *Prior Storage1*
Serve the *Content Segment*; update *Utility1*
**If** *Utility1* is greater than *THRESHOLD*
promote the *Content Segment* to *Prior Storage2* replacing a content segment (if needed).
**Else** Prior storage the *Content Segment* in *Prior Storage1* evicting a content segment based on *Utility1*.
**Initialize/update** utility of all contents segments in Prior Storage.

Whenever it enters the system for the first time. This is particularly beneficial when the request arrival rate is very high. The minimum time for which a content segment is kept in *Prior Storage1* is the maximum of either the playback time of the content segment or the playback time of the current request.

**2.3. Operation Functions in the Optical Network Unit**

Figure 2 shows the new traffic classifier and the IPTV controller components in the proposed ONU architecture. The traffic classifier has four modules, include CoS classifier, Routing table, Packet forward and Ingress rule, which enables the ONU to classify the user traffic based on different parameters. The ONU with an IPTV controller delivers the live IPTV traffic to the end user by handling the requests as either inter traffic or intra traffic (local traffic). Inter traffic is the traffic between the ONU and OLT, such as video content streams, and intra traffic is the traffic between an ONU and users without sending information to the OLT via the feeder fibre, such as control signals.

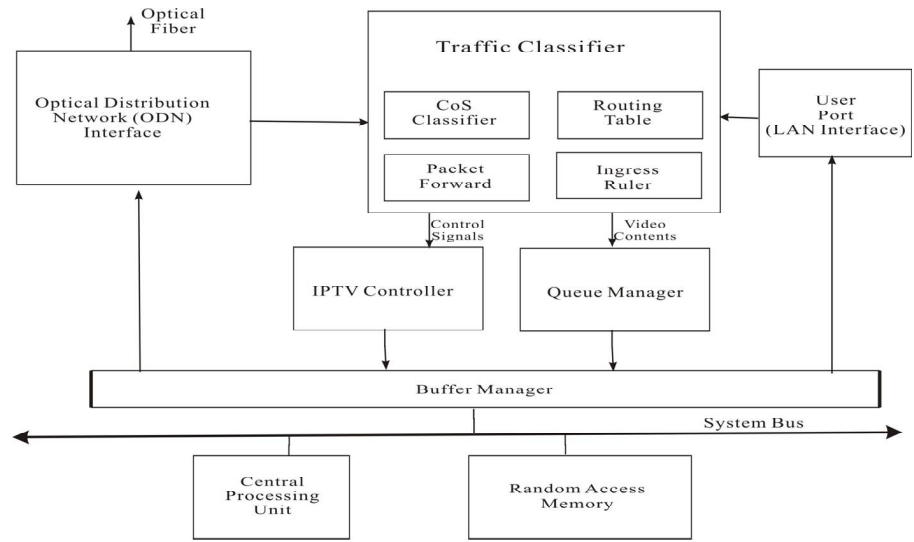

Figure 2. Proposed Optical Network Unit Architecture

RTSP (IPTV channel) requests will be forwarded to the IPTV controller. The IPTV controller consists of a packet analyser and control functions. It processes the requests up to the application layer in the OSI model. If the requested channel is already multicast by the OLT and the Channel Logical Link Identifier (CLLID) exists in the ONU table, the ONU handles control signals in the request as local traffic. On the other hand, if the requested channel is not broadcast by the OLT or the CLLID does not exist in the ONU table, the request will be sent to the OLT, and the OLT adds the desired CLLID to the ONU table. The main function of the IPTV controller is handling control signals as a part of IPTV requests in local traffic. Moreover, to manage the multicast operation and ONU table, the IPTV controller of ONU cooperates with the IPTV engine of the OLT. The functions in the IPTV controller of ONU are mandatory to handle the operations. The extract function is extracted the channel name form the packet. The check function is to determine whether the channel name exists in the routing table at ONU. The Retrieve_metadata (channel_name,the user_MAC) function to send the metadata to the user. The Send_metadata (channel_name, ONU_LLID) is sent the metadata to the ONU. The Add_table (CLLID, name, user_MAC) function is executed to add a record to the ONU routing table. When the user requests to leave the channel, the Stop (channel_name, user_MAC) and Remove_table (CLLID, channel_name, user_MAC) functions will be executed to stop broadcasting the channel to the user and keep the ONU routing table up to date. The authentication function is responded to Add function with Authentication (channel_name, user_info) function. If the user is without the authorization to receive the channel data, sending the data will be stopped by executing the functions Stop (channel_name, user_MAC) and Remove_table (CLLID, channel_name, user_MAC).

## 2.4. Operation Functions in the Optical Line Terminal

Figure 3 shows the traffic classifier and the IPTV engine components in the proposed OLT architecture. The traffic classifier is used to classify the IPTV traffic in the downstream/ upstream direction. The IPTV traffic classifier in the OLT redirects the data stream from the Video Contents Head Office to the IPTV engine for further processing. Moreover, the requests are passed by the ONU IPTV controller forward to the IPTV engine by the traffic classifier for handling the request with proper operations. The IPTV engine consists of the firmware, the microprocessor unit (MPU), the buffer, and the memory storage, the packet processing engine. The packet processing engine processes the packet up to the application layer in the OSI model. The firmware contains the functions and algorithms to handle the requests and updates the ONU/OLT tables.

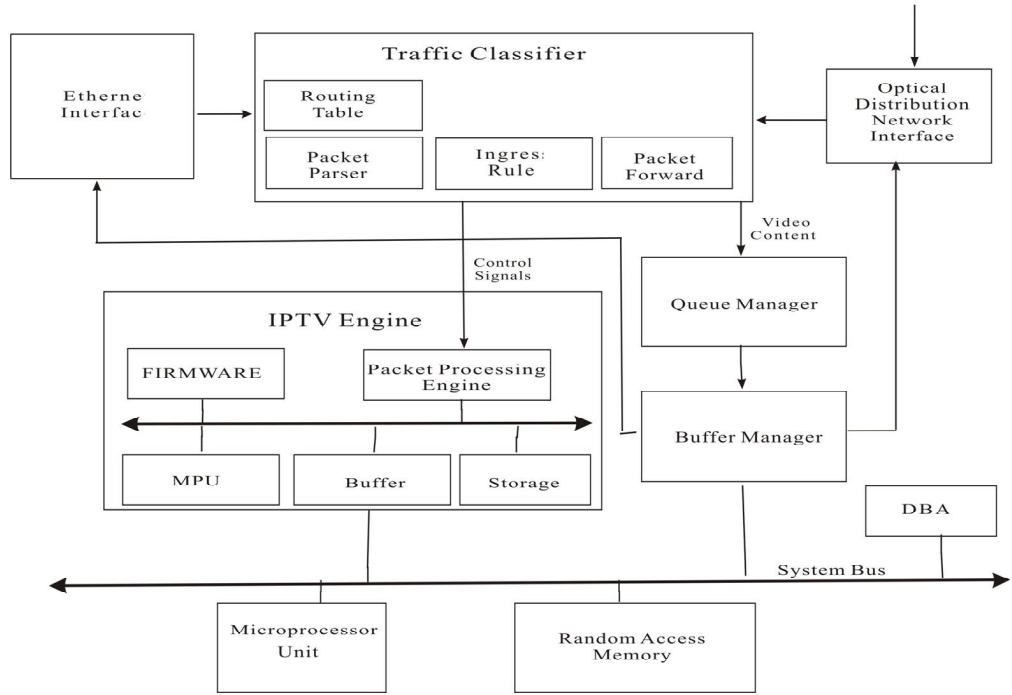

Figure 3. Proposed Optical Line Terminal Architecture

When an ONU forwards a request, the OLT will retrieve the Request_LLID (channel_name) function to check the CLLID to determine whether the requested channel already exists or not, if the request function should assign a unique LLID by executing the Add_LLID (channel_name, CLLID) function. The OLT gets the Add_ONU (channel_name, ONU_LLID) to add a record to the ONU table and update its information. After updating the ONU and OLT table, the OLT begins to send channel streaming by putting the CLLID in the frames. The OLT routing table controls starting and stopping the data channel multicasting. When there is no user request for the channel in the ONU routing table, the ONU requests the OLT to remove the ONU LLID from the OLT routing table by executing the Delete (channel_name, ONU LLID) function in the IPTV engine. If there is no ONU LLID in the record for the channel, the OLT stops multicasting the data by executing the Stop (CLLID) function.

## 4. CONCLUSIONS

The major concern for designing IPTV service system is storage and bandwidth limitation. An efficient approach to address this concern is to deploy distributed content prior stores and to prior store popular content that are likely to be used in the near future. In this paper, we proposed a dual level prior storage algorithm which adaptively prior storage with bi-level control. It incorporates well the flattened head popularity characteristics of IPTV on demand content streaming service. According to our trace-driven simulation experiments, the proposed bi-level control caching algorithm achieves very high hit ratio which is very superior to the traditional replacement policy. Furthermore, if prior storage management may combine it with simple prior store placement policy to address heavy tailed unpopular objects, storage size is relatively large. The problem of prior store management is to design a good replacement policy that maximizes the hit ratio measured over a very long trace. It is subject to the important practical constraints of minimizing the computational and space overhead involved in implementing the policy. But, if a prior store management algorithm has parameters to be tuned, such as, the video content prior storing strategy has two threshold values for control, i.e., upper and lower limits, $T_H$ and $T_L$ ($0 < T_L < T_H < N$) respectively. In the proposed algorithm and requires a prior knowledge on system behaviour, it cannot guarantee the optimal performance uniformly well across different workloads and storage memory sizes. We propose a QoS test approach at optical access networks and integrate the prior storage scheme management methodology for the ISP/NSP to verify whether the optical subscriber lines of access network provide sufficient network resource. We define the QoS parameters and illustrate the integrated architecture of the test mechanism with operation functions of ONU and OLT into the optical access networks. Finally, execute the verification test.

Therefore, as further works, user-defined parameters need to be adapted online and on-the-fly fashion.

**Author**

**Tony Tsang**

Tony Tsang (M'2000) received the BEng degree in Electronics & Electrical Engineering with First Class Honours in U.K., in 1992. He received the Ph.D from the La Trobe University (Australia) in 2000. He was awarded the La Trobe University Post-graduation Scholarship in 1998. Prior to joining the Hong Kong Polytechnic University, Dr. Tsang earned several years 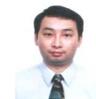 of teaching and researching experience in the Department of Computer Science and Computer Engineering, La Trobe University. He works in Hong Kong Polytechnic University as Lecturer since 2001. He works in Chu Hai College in 2015. He has numerous publications in international journals and conferences and is a technical reviewer for several international journals and conferences. His research interests include mobile computing, networking, protocol engineering and formal methods. Dr. Tsang is a member of the ACM and the IEEE.